\documentclass[letterpaper,journal]{IEEEtran}

\usepackage{graphicx}
\usepackage{mathrsfs}
\usepackage{amsfonts}
\usepackage{amsmath}
\usepackage{amssymb}
\usepackage{amsthm}
\usepackage{array}
\usepackage{url}
\usepackage{xcolor}
\usepackage{lipsum}
\usepackage{comment} 
\usepackage{subfig} 
\usepackage{multirow} 
\usepackage{floatrow}
\usepackage{mdframed}
\usepackage{floatpag}

\newcommand{\change}[1]{{\color{black}#1}}

\theoremstyle{remark}
\newtheorem{remark}{Remark}

\def\BibTeX{{\rm B\kern-.05em{\sc i\kern-.025em b}\kern-.08em
    T\kern-.1667em\lower.7ex\hbox{E}\kern-.125emX}}

\begin{document}

\title{The Twin-in-the-Loop approach\\for vehicle dynamics control}
\author{Federico~Dettù, 
Simone~Formentin and~Sergio~Matteo~Savaresi%,~\IEEEmembership{Life~Fellow,~IEEE}% <-this % stops a space
	\thanks{This research was partially funded by company VI-Grade SrL, whose technical support is here gratefully acknowledged.}
	\thanks{All the authors are with Dipartimento di Elettronica, Informazione e Bioingegneria, Politecnico di Milano, Piazza Leonardo da Vinci 32, 20133, Milano, Italy (e-mail: federico.dettu@polimi.it, simone.formentin@polimi.it, sergio.savaresi@polimi.it).}
}

%\IEEEoverridecommandlockouts
%\IEEEpubid{\makebox[\columnwidth]{978-1-5386-5541-2/18/\$31.00~\copyright2018 IEEE \hfill}
%	\hspace{\columnsep}\makebox[\columnwidth]{ }}
%\maketitle
%\IEEEpubidadjcol

%\end{minipage}
\maketitle
\begin{figure}[!ht]
	\begin{mdframed}[linewidth=0.5pt,linecolor=black]
		© 2023 IEEE. This article has been accepted for publication in IEEE/ASME Transaction on Mechatronics. This is the author's version which has not been fully edited and
		content may change prior to final publication. Personal use of this material is permitted. Citation information: DOI 10.1109/TMECH.2023.3292503.
	    Personal use is permitted, but republication/redistribution requires IEEE permission.
     	See https://www.ieee.org/publications/rights/index.html for more information. 
	\end{mdframed}
\end{figure}

\begin{abstract}
In vehicle dynamics control, engineering a suitable regulator is a long and costly process. The starting point is usually the design of a nominal controller based on a simple control-oriented model and its testing on a full-fledged simulator. Then, many driving hours are required during the End-of-Line (EoL) tuning phase to calibrate the controller for the physical vehicle. Given the recent technological advances, we consider in this paper the pioneering perspective where the simulator can be run \textit{on-board} in the electronic control unit, to calculate the nominal control action in real-time. In this way, it can be shown that, in the EoL phase, we only need to tune a simple compensator of the mismatch between the expected and the measured outputs. The resulting approach not only exploits the already available simulator and nominal controller and significantly simplifies the design process, but also outperforms the state-of-the-art in terms of tracking accuracy and robustness within a challenging active braking control case study. 
\end{abstract}

\section{Introduction}
\label{Section:Introduction}
The use of Digital Twins (DTs) - combining software and physical connections to produce a faithful virtual replica of a given system - is revolutionizing state-of-the-art technological solutions in different fields \cite{bhatti_2021}. 

In the automotive world, DTs are highly exploited for on-line monitoring and prognostics of vehicle components \cite{bhatti_2021}. Moreover, vehicle dynamics simulators - as faithful replicas of the system - are widely employed at the mechanical design level, \textit{i.e.}, when selecting physical components or when assessing the differences among structural choices \cite{kutluay_2014}. Instead, vehicle dynamics controls are still based on simple control-oriented models, usually capturing the key features of a single maneuver \cite{lucchini_2020}. Indeed, this yields well known issues when implementing the controller on the real platform, having to deal with many unmodeled dynamics. In the industrial practice, this issue is overcome by finely adjusting the controller parameters, during the so-called End-of-Line (EoL) tuning process \cite{tanelli_2011}. Nonetheless, the latter might be a time consuming and costly procedure, especially when considering complex industrial regulators.

In this paper, we take advantage of the most recent technological advances to show how a high-fidelity vehicle simulator can be used \textit{in real-time} and \textit{directly embedded} into the Electronic Control Unit (ECU) of the vehicle to enhance the closed-loop performance and significantly simplify the EoL tuning phase. We will denote the resulting architecture \textit{Twin-in-the-loop} (TiL) \textit{control} hereafter. 
Doing so, the nominal control action could be provided as the one computed on the system simulator, whereas only dynamics to be controlled is the mismatch between the vehicle and the simulator, which could be handled by a simpler compensator. 
We remark that, even if the possibility of running a full-fledged vehicle simulator on-board may sound far in the future, it rather fits well into the present context of an ever-increasing on-board and cloud-based \cite{bhatti_2021} computing power, driven by the necessity of processing signals coming from visual sensors, such as LiDARs. Indeed, this novel approach calls for a tuning procedure for the additional compensator, which will be also treated in this paper.

TiL shares some common feature with \textit{internal model control} (IMC) \cite{garcia_1982}. In IMC, an approximated model of the plant can be used to generate the control action necessary to obtain a desired output. In fact, if the model is invertible (\emph{e.g.} in case of minimum phase transfer functions), the nominal control action is easily obtained; if not, the model has to be decomposed into its invertible and non-invertible parts, \emph{e.g.} by adding high-frequency poles to the non-causal model inverse. IMC has been recently applied for spark ignition engines control \cite{ossareh_2021}, \cite{wu_2019}, but the applications to vehicle dynamics are not frequent in the literature, with one notable exception for the yaw-rate controller in \cite{canale_2008}. \\ An important problem in IMC - that can be read as a possible reason for its low popularity in spite of other approaches, like, \emph{e.g.}, Model Predictive Control - lies in the strong requirement on the plant to be invertible or at least decomposable with an invertible part; when the plant cannot be inverted, nor its equation can be written in closed-form, IMC cannot be applied. This is the case of commercial driving simulators, which are often black-box objects, and can only be used to obtain a certain output based a specified set of inputs. TiL solves this invertibility problem for this class of models, as we show later on in this manuscript.

In order to show the potential of TiL architecture as compared to traditional vehicle controls end-of-line tuning, we consider a well-known and challenging safety-critical problem, namely \textit{active braking control} \cite{savaresi_2010}. \textit{Active braking control} - or anti-lock braking system (ABS) - is a key piece of software in modern vehicles; its main purpose is to keep the longitudinal slip around the friction peak to maximize braking performance and thus avoiding dangerous wheel lock. \\
More specifically, we consider a high-performance sport car as the vehicle to be controlled, the latter being an interesting proving ground for paradigm shifting technologies, due to the extremely challenging and highly dynamic driving conditions. Model predictive control (MPC) has been recently employed in braking control \cite{riva_2022,tavernini_2019,basrah_2017}, in order to explicitly account for actuator dynamics and constraints in the control law - which is impossible in simpler controls, such as PID \cite{johansen_2003}. For these reasons, we consider an MPC to build our proof-of-concept: specifically, we select the approach proposed in \cite{riva_2022}, where a specific design for high-performance vehicles is proposed. This choice is indeed not a strong assumption, as TiL can be in principle used to enhance any control algorithm and to simplify its implementation; we will discuss this in detail later on.
%For fair benchmarking, since model predictive control (MPC) has been recently applied to such a problem obtaining unparalleled performance \cite{riva_2022}, we build up our proof-of-concept starting from that research. \\
Finally, we wish to remark here that recent work has already shown the potential of such a technology for accurate vehicle state estimation \cite{riva_2022_SIL}; a recent development \cite{delcaro_2023} also shows how to deal with the dimensionality issues when calibrating TiL estimators. Indeed, to the best of our knowledge, this is the first time that the TiL paradigm is applied to control systems.

The remainder of the paper is as follows. Section \ref{Section:Problem_statement} formally states the problem, while Section \ref{Section:case_study} discusses the active braking case study. Section \ref{Section:silc_braking} applies TiL control to the proposed case study, and Section \ref{Section:ControllerTuning} shows how the algorithm parameters can be calibrated starting from a set of closed-loop experiments. Section \ref{Section:simulation_results} compares the TiL solution to the benchmark; finally, Section \ref{Section:HIL} provides an Hardware-in-the-Loop real-time validation of the proposed controller on an off-the-shelf ECU, in order to prove real-time feasibility of the same. The paper is ended by some concluding remarks.

%%%%%%%%%%%%%%%%%%%%%%%%%%%%%%%%%%%%%%%%%%%%%%%%%%%%%%%%%%%%%%%%%%%%%%%%%%%%%%%%%%%%%%%%%%%%%
\section{Problem statement}
\label{Section:Problem_statement}
\begin{figure}[h]
	\centering
	\includegraphics[width=0.8 \columnwidth]{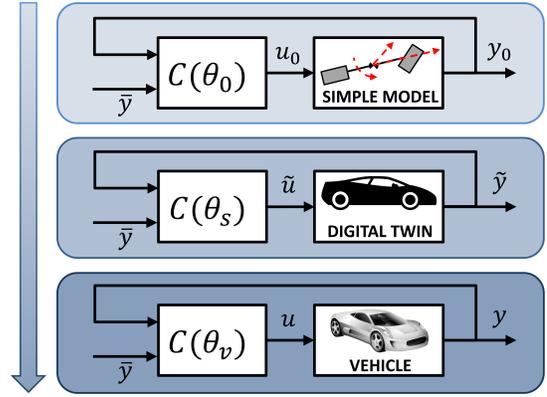}
	\caption{Traditional vehicle controls development: the controller is first designed based on a simple control-oriented model, then it is tested on a full-fledged simulator and finally it is fine tuned on the physical vehicle.}
	\label{fig:control_development_framework}
\end{figure}

The traditional framework for the development of vehicle dynamics control systems usually follows the three-step procedure depicted in Fig. \ref{fig:control_development_framework}, in which we have considered the case of reference following without loss of generality.

Specifically, the full design procedure is as follows.
\begin{enumerate}
	\item A controller $C(\theta)$, fully described by a set of parameters $\theta$, is first designed based upon a control oriented model, capturing the main dynamics of interest - e.g. a quarter-car model in case of active braking control \cite{riva_2022}. From now on, we denote the set of parameters selected using the simple model as $\theta_0$.
	\item The controller is then tested on a high-fidelity multibody simulator (a digital twin), usually accounting for unmodeled behaviours and nonlinearities. The initial values $\theta_0$ are then adjusted until acceptable performance is reached, leading to a new optimal selection $\theta_s$
	% Given a simulator model $\tilde{\mathcal{M}}$, this reads
	%\begin{equation}
	%	\tilde{y}=\mathcal{\tilde{M}}\left(\tilde{u}\right),\ \tilde{u}\in \mathbb{R}^p, \tilde{y}\in \mathbb{R}^{m},
	%\end{equation}
	%where $\tilde{u}$ is generated in closed-loop.
	\item Finally, the controller is implemented on an Electronic Control Unit (ECU) and tested on the physical vehicle along a few driving hours. This step usually requires a final refinement of the parameters, yielding the new vector $\theta_v$, due, \textit{e.g.}, to neglected phenomena or measurement noises. 
\end{enumerate}

\begin{figure}[h]
	\centering
	\subfloat[TiL-C control architecture.\label{fig:sil_c_scheme}]{\includegraphics[width=0.8 \columnwidth]{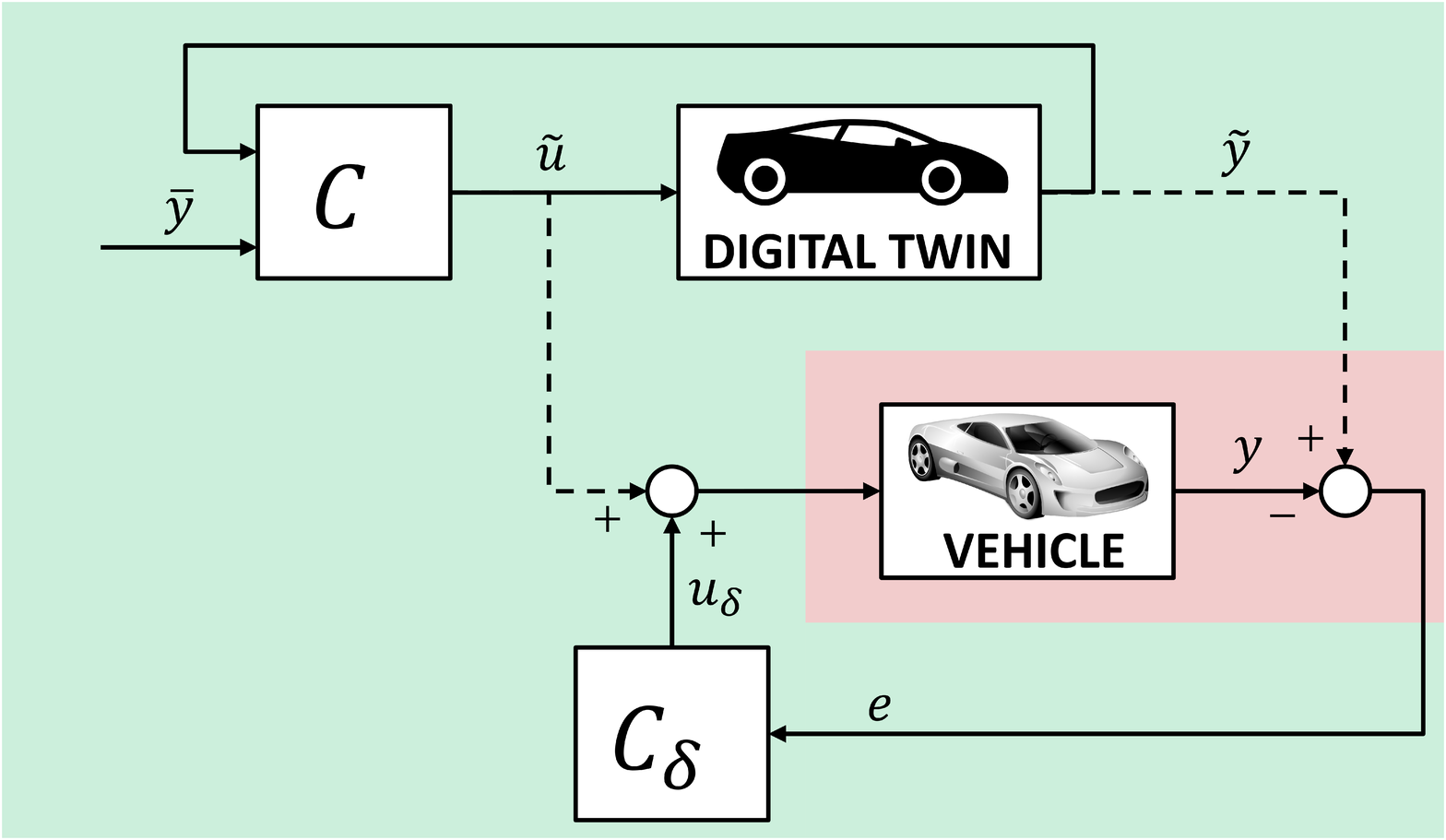}}
	\hfill
	\subfloat[TiL-C implementation.\label{fig:sil_c_implementation_v2}]{\includegraphics[width=0.8 \columnwidth]{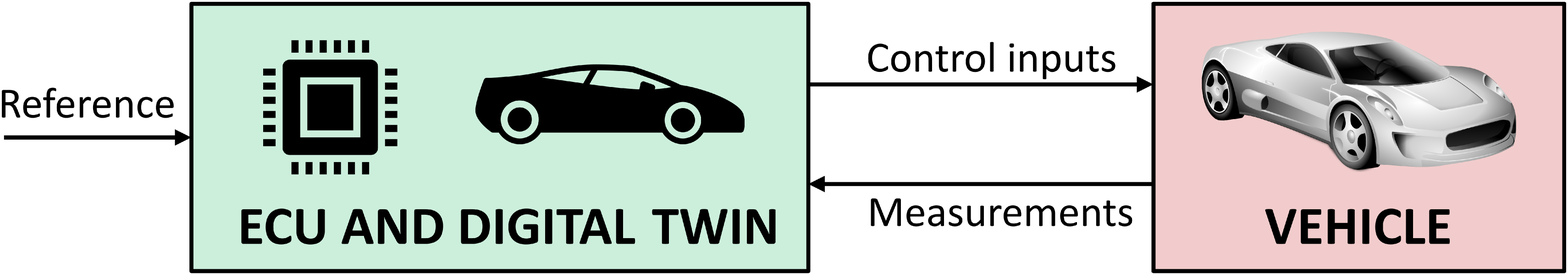}}
	\caption{TiL-C architecture (top) and its on-board implementation (bottom). The red areas denote the physical vehicle, whereas the green areas depict the software elements, namely the simulator and the controllers, running on an ECU.}
\end{figure}

The key observation behind the idea of this work is that, once $C\left(\theta_s\right)$ has been tuned and tested on the simulator, an \textit{ideal} control action $\tilde{u}$ and the corresponding output $\tilde{y}$ become available. The latter point opens the possibility to a \textit{quantum leap} in the field of vehicle controls: if the simulator can be run in real-time on the ECU, the ideal $\tilde{u}$ can be used as a nominal control action with no additional computations. If the simulator and the vehicle dynamics coincide, such an input could be directly applied to the real system. Indeed, if the simulator and the vehicle differ in some way, a second control loop needs to be designed accounting for the mismatch between $\tilde{y}$ and the measured $y$. Let us denote the additional compensator as $C_{\delta}$. Doing so, the control action on the physical vehicle becomes equal to $u_{\delta}+\tilde{u}$.
We denote the above described control architecture as \textit{Twin-in-the-Loop} Control (TiL-C), and depict its schematics in Fig. \ref{fig:sil_c_scheme}.

The proposed architecture shows a set of interesting features as compared to the standard practice:
\begin{itemize}
	\item if the simulator is a faithful replica of the vehicle, most of the system nonlinearity and complexity is managed by $C$, and \textit{the nominal control action does most of the job}. The compensator $C_{\delta}$ would thus play the role of the controller of a linearized system in the neighborhood of an operating point;
	\item since $C$ is operated on the simulator, it has access to all its states. This opens up the possibility of designing \textit{state-feedback approaches} (for the nominal control action) even when the state is not (fully) available without the need of designing suitable observers;
	\item the EoL tuning procedure of Fig. \ref{fig:control_development_framework} cannot be avoided, as $C_{\delta}$ needs to be tuned based on the mismatch between the vehicle and the simulator. However, \textit{the design might be largely simplified}, as $C_{\delta}$ can be selected as a simple controller with few parameters even when $C$ is a complex (possibly optimization-based) controller. 
	\item the TiL-C approach can in principle be \textit{generalized} to any vehicle dynamics problem, classical examples being longitudinal, lateral and vertical dynamics control. In fact, many of these tasks can be formulated as reference tracking problems (see, \emph{e.g.}, the yaw-rate tracking problem in \cite{lucchini_2020} or \cite{canale_2008}), thus the control scheme is conceptually the same of Fig. \ref{fig:sil_c_scheme}. 
\end{itemize}
Indeed, tuning the Digital Twin of scheme in Fig. \ref{fig:sil_c_scheme} requires proper calibration. In what follows, we will assume that such calibration has been already performed by the car manufacturer according to standard practices \cite{kutluay_2014}, so this topic is left out of the scopes of the paper.
\begin{remark}
Note that the TiL scheme presented in this research might resemble an Hardware-in-the-Loop (HiL) architecture running on-board at real-time, where a faithful plant replica runs in-silico. However, there are important differences between HiL and TiL. On the one hand, the former is usually a test-bench for controller implementation \cite{fathy_2006}, where some piece of hardware (\emph{e.g.} an actuator, or an Electronic Control Unit) is inserted in the loop. On the other hand, in TiL architectures, the commands generated from the simulator are directly used to control the vehicle.
\end{remark}
\section{The case study: active braking control}
\label{Section:case_study}

\begin{figure}[h]
	\centering
	\includegraphics[width=0.8 \columnwidth]{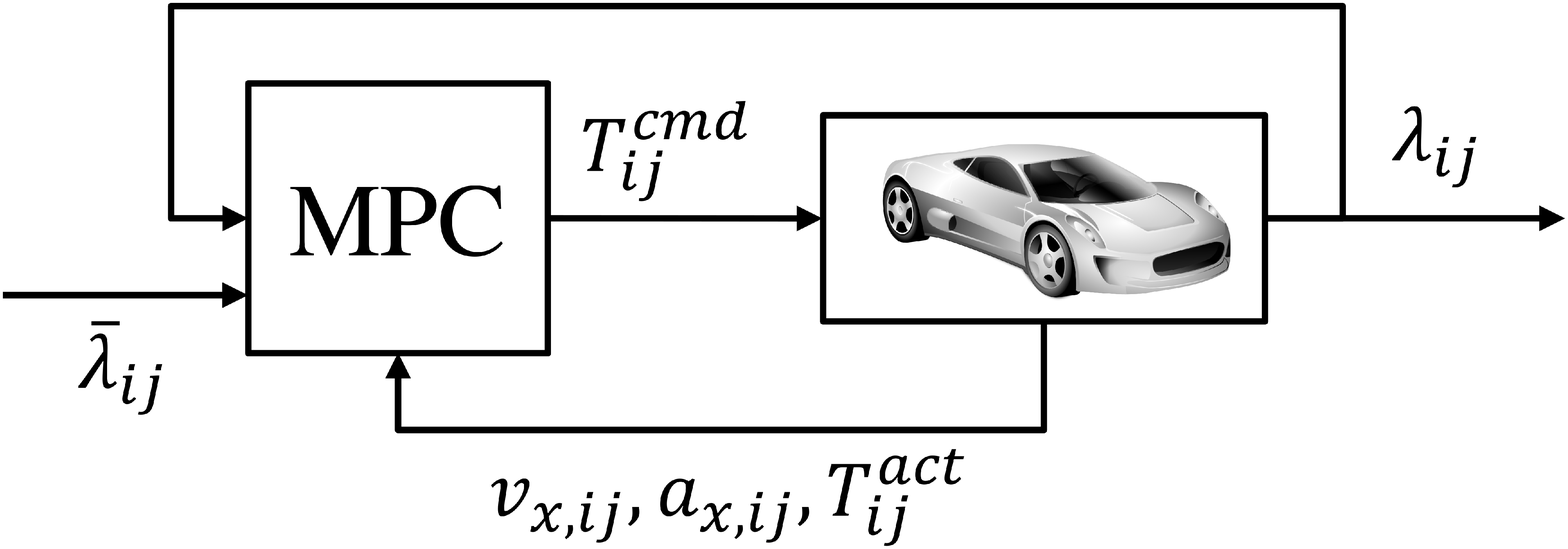}
	\caption{MPC-based braking control \cite{riva_2022}. The torques are computed independently for each wheel, then $ij=fl,fr,rl,rr$.}
	\label{fig:mpc_braking}
\end{figure}

As a case study to illustrate the potential of the proposed approach, let us consider the problem of longitudinal dynamics control during braking, \textit{i.e.}, the design of an Anti-lock Braking System (ABS) for a high-performance car. In particular, we focus on the control of the wheel slip $\lambda$ aimed to track a desired behaviour $\bar{\lambda}$ so that a certain braking force can be guaranteed, and the vehicle can stop without wheel locking. In this seminal study, we will consider two instances of the same simulator (as detailed next) to play the role of the digital twin and the real car. The problem of selecting $\bar{\lambda}$ is instead out of the scopes of this work and will not be discussed here.
Also, note that an ABS controller is generally independent on the outside road events, like traffic conditions: the ABS is activated an deactivated solely based on a request by the driver, which can stop its action at any time by releasing the brake pedal.
We wish to stress that the main goal of this research is not to design a full braking control algorithm \textit{per se}, but to simplify the EoL tuning problem by relying on the availability of the simulator on-board. 
As the nominal controller $C$ in Fig. \ref{fig:sil_c_scheme}, we thus start from the existing, state-of-the-art, MPC approach in \cite{riva_2022}. The control action is computed independently for each wheel, thus $ij=fl,fr,rl,rr$, as showed in Fig. \ref{fig:mpc_braking}. In the figure, we also highlight the feedback measurements necessary for the MPC, which are the wheel ground contact point velocity $v_{x,ij}$, the actuated torque $T^{act}_{ij}$ and the wheel longitudinal acceleration $a_{x,ij}$.

The wheel slip $\lambda$ at corner $ij$ is defined as 
\begin{equation}
	\lambda_{ij}=\dfrac{v_{x,ij}-\omega_{ij}R_{ij}}{\max\left\{v_{x,ij},\omega_{ij}R_{ij}\right\}},
	\label{eq:slip_eq}
\end{equation}
where $\omega_{ij}$ is the wheel angular rate and $R_{ij}$ is the wheel radius \cite{savaresi_2010}. Note that the slip in Eq. \ref{eq:slip_eq} is always positive in case of braking, as the numerator is always $<0$, since the wheel decelerates.\\

We consider straight braking, \textit{i.e.}, we assume almost zero camber and sideslip wheel angles. Such an assumption does not affect the scope and outcomes of the research; it is in fact well known that such variations affect the friction analogously to the vertical wheel loads, and can be neglected by suitably scheduling the slip references \cite{savaresi_2010}.

\subsection{The vehicle and the simulator}
\label{Section:VehicleModel}
Both the vehicle - a sport car - and its digital twins are here modeled in the VI-Grade CarRealTime (CRT) simulation environment \cite{vigrade_2022}. The CRT simulator include a 6 degree-of-freedom (DOF) object for the chassis, a 1-DOF model for the suspensions and a 1-DOF model for the wheel dynamics. A model of the electro-hydraulic brake (EHB) is also added, so as to take into account realistic actuation limits. The most critical vehicle parameters influencing the dynamics during braking \cite{savaresi_2010} are given in Table \ref{Tab:vehicle_parameters}. 
\begin{table}[thpb]
	\caption{Braking-related parameters for the considered car.}
	\label{Tab:vehicle_parameters}
	\begin{center}	
		\begin{tabular}{|c|c|c|c|}
			\hline 
			$M_{tot}\ [kg]$ & $R^{nom}_f\ [m]$ & $R^{nom}_r\ [m]$ & $J^{nom}_f\ [kg \cdot m^2]$ \\
			\hline	
			$1612$ & $0.33$ & $0.35$ & $1.49$ \\
			\hline
			\hline
			$J^{nom}_r\ [kg \cdot m^2]$& $l_f\ [m]$ & $l_r\ [m]$ & $h\ [m]$ \\
			\hline
			$2.25$ & $1.57$ & $1.03$ & $0.46$\\
			\hline
		\end{tabular}
	\end{center}
\end{table}

$M_{tot}$ represents the total vehicle mass, encompassing sprung and unsprung masses, whereas $R^{nom}_f$, $R^{nom}_r$ represent the nominal wheel radius, for front and rear wheels respectively; note that the high-fidelity model also accounts for radius variations due to, \textit{e.g.}, increased wheel loads. $J^{nom}_{f}$ and $J^{nom}_{r}$ are the nominal spin inertia of front and rear wheels. $l_f$ and $l_r$ are the distances between the projection of the vehicle center-of-gravity (COG) on the ground and the front and the front and rear axles, respectively. $h$ is the COG height from the ground.

Formally, the control inputs $u$ and the set of driver commands $\xi$ are
\begin{equation}
	\begin{split}
		u=&\begin{bmatrix}
			T^{cmd}_{fl} & T^{cmd}_{fr} & T^{cmd}_{rl} & T^{cmd}_{rr}
		\end{bmatrix}^t, \\
		{\xi}=&\begin{bmatrix} {\phi}_{throttle} & {\phi}_{brake} & {\phi}_{steer} & {\phi}_{gear} \end{bmatrix}^t
	\end{split}
\end{equation}
where ${T}^{cmd}_{ij}$ represents a torque command at wheel $ij$,  $\phi_{throttle},\ \phi_{brake},\ \phi_{steer}$ are respectively throttle, brake and steer driver request, while $\phi_{gear}$ is the inserted transmission gear. Indeed, when the braking controller is active, the driver brake request is bypassed. Due to the straight braking assumption, throttle and steer commands are negligible. The variables $\xi$ simultaneously act on the simulator and on the vehicle, so the driver should be considered as an exogenous disturbance in the framework of Fig. \ref{fig:mpc_braking}.

\begin{figure}[h]
	\centering
	\includegraphics[width=0.7 \columnwidth]{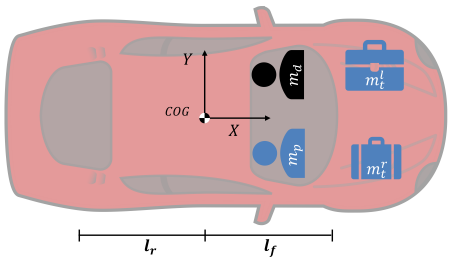}
	\caption{The considered vehicle with modeled (black) and unmodeled (light blue) masses.}
	\label{fig:vehicle_model}
\end{figure}

In order to simulate a realistic difference between the vehicle and its digital twin, 
we will consider, in three different case studies, the effect of realistic measurement noises, the effect of a different mass distribution and of a variation of the tire-road force diagram.
Specifically, since high performance cars are usually characterized by two front seats and a front trunk, we add a second concentrated mass on the passenger seat, and two unbalanced masses on the front trunk. A representation of the unmodeled loads is provided in Fig. \ref{fig:vehicle_model}, where it becomes clear how the presence of unmodelled masses changes the ratio between the distance of COG and front wheels $l_f$ and that of COG and rear wheels $l_r$, thus varying the front/rear load tranfer.
\begin{figure}[h]
	\centering
	\includegraphics[width=0.9 \columnwidth]{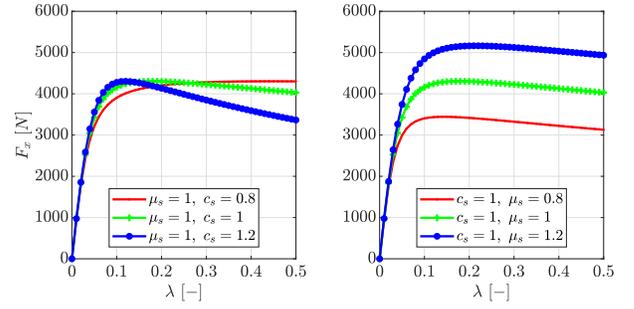}
	\caption{Front wheel longitudinal force versus slip, in case of nominal vertical tire force. Left and right plots depict variations of the shape factor and peak friction, respectively.}
	\label{fig:pacejka_2}
\end{figure}
The values of the additional masses are as in Table \ref{Tab:vehicle_masses}.
\begin{table}[thpb]
	\caption{Additional masses included in the vehicle model. Such masses are modeled as concentrated loads.}
	\label{Tab:vehicle_masses}
	\begin{center}	
		\begin{tabular}{|c|c|c|c|}
			\hline 
			$m_d\ [kg]$ & $m_p\ [kg]$ & $m^l_t\ [kg]$ & $m^r_t\ [kg]$\\
			\hline	
			$75$ & $80$ & $90$ & $30$ \\ 
			\hline	
		\end{tabular}
	\end{center}
\end{table} \\
\textit{VI-CarRealTime} models tire-road interaction via a set of Pacejka formulas, governing the longitudinal and lateral force generation at the tire - as well as more complex phenomena, such as overturning moment. In order to further test the control robustness, we consider uncertainty in the longitudinal force generation, specifically by simulating a multiplicative mismatch $\mu_s$ on the peak friction and $c_s$ on the shape factor. Figure \ref{fig:pacejka_2} displays the nominal behaviour ($\mu_s=c_s=1$), and different versions of the same model with perturbations of the two parameters.
\subsection{Model predictive controller}
\label{Section:MPC}
The MPC in \cite{riva_2022} is here considered as a nominal controller and is assumed to be designed before-hand, in the pre-development stage. Such a controller employs a wheel-specific predictive model of the slip dynamics
\begin{equation}
	\begin{split}
		\dot{\lambda}^{mpc}_{ij}&=\dfrac{1-\lambda^{mpc}_{ij}}{v_{x,ij}} a_{x,ij}-\dfrac{R^{nom}_{ij}}{J^{nom}_{ij} v_{x,ij}}T^{act}_{ij}, \\
	\end{split}
	\label{eq:predictive_slip}
\end{equation}
where $\lambda^{mpc}_{ij}$ represents the prediction of the slip at each wheel. Parameters $R^{nom}_{ij}$, $J^{nom}_{ij}$ are taken from Table \ref{Tab:vehicle_parameters}. 
The actuator dynamics is a second order system with the same model and parameters of \cite{riva_2022}.
The model \eqref{eq:predictive_slip} is used in the MPC under the assumption of constant wheel speed and acceleration during the prediction horizon $T_{p}$, namely
\begin{equation}
	\begin{split}
		a_{x,ij}\left(t_0+t\right)&=a_{x,ij}\left(t_0\right),\ \forall t \in\left[t_0,t_0+T_p\right] \\
		v_{x,ij}\left(t_0+t\right)&=v_{x,ij}\left(t_0\right),\ \forall t \in\left[t_0,t_0+T_p\right].
	\end{split} 
\end{equation}
Where $t_0$ is the current time instant, and $t$ is the independent time variable.
Under the assumptions above - motivated by the different time scales between slip and chassis longitudinal dynamics \cite{savaresi_2010} - the model in Eq. \eqref{eq:predictive_slip} becomes linear and time-invariant (LTI). Said LTI model can be then discretized and written in velocity form \cite{magni_2014} - \textit{i.e.}, transforming states and inputs into their instantaneous variations; the slip tracking error is then further introduced as a state. At this point, an integral action can be easily implemented in the MPC: such formulation has been shown to be robust to constant disturbances, guaranteeing zero steady-state error. \\
Five predictive steps are considered in the computation of the optimal control law, which is computed implicitly; with regards to other details regarding the optimal control formulation, we refer the reader to \cite{riva_2022}.

\subsection{Sensor model}
\label{Section:noise_model}
Any controller acting on the system described in Section \ref{Section:VehicleModel} is based upon sensor measurements. For a realistic case study, we cannot neglect the effect of the noise model on the performance. In particular, we introduce the noise affecting speed, acceleration and slip as detailed next. No noise is added to the braking torque, assuming it is fully controllable and known.
%On the other hand, slip cannot be directly measured, and is instead estimated resorting to equation

\textbf{Acceleration.} Longitudinal acceleration measurements at each wheel ($a_{x,ij}$) are seldom available on production vehicles. Chassis acceleration is then employed when considering straight braking \cite{riva_2022}. Such signal is obtained through a Inertial Measurement Unit (IMU), usually affected by high-frequency noise \cite{campos_2022}. We thus consider random Gaussian noise in the acceleration measurement $a_x^n$
\begin{equation}
	a_x^n=a_x+n_a,\ n_a \sim WN(0,\sigma_a^2),
	\label{eq:noise_acceleration}
\end{equation}
where $n_a\sim WN\left(\mu,\sigma^2\right)$ denotes a Gaussian noise with expected value $\mu$ and variance $\sigma^2$.

\textbf{Chassis speed.} The COG-referenced longitudinal speed $v_x$ cannot be directly measured without high accuracy Global Positioning System (GPS) sensors, then it is usually estimated through state observers \cite{hashemi_2016}, and reported at each wheel via kinematic relations. Hence, we consider low-pass filtered version $n_{lp}$ of a white noise $n_{v_x}$ on the speed measurement $v_x^n$, to mimic the state observer dynamics
\begin{equation}
	v^{n}_{x}=v_{x}+n_{lp}\left(n_{v_x}\right),\ n_{v_x} \sim WN(0,\sigma_{v_x}^2),
	\label{eq:noise_speed}
\end{equation}
where $\sigma^2_{v_x}$ is the speed white noise variance.\\

\textbf{Wheel speed.} The angular rates are measured through incremental encoders: such sensors are well known to be affected by periodic noise \cite{panzani_2012}, mostly due to unavoidable geometrical or misalignment errors in the sensor structure. The amplitude of such a noise increases depending on the rotational speed itself. We thus include a speed-scheduled sinusoidal error term $\omega^n_{ij}$
\begin{equation}
	\omega^n_{ij}=\omega_{ij}+A_{\omega}\sin \left(\omega_{ij} t\right),\ A_{\omega}=\omega^n_0+k_\omega \omega_{ij},
	\label{eq:noise_wheelrate}
\end{equation}
where $\omega_0^n$ and $k_\omega$ are tunable parameters.\\

\begin{figure}[h]
	\centering
	\includegraphics[width=0.9 \columnwidth]{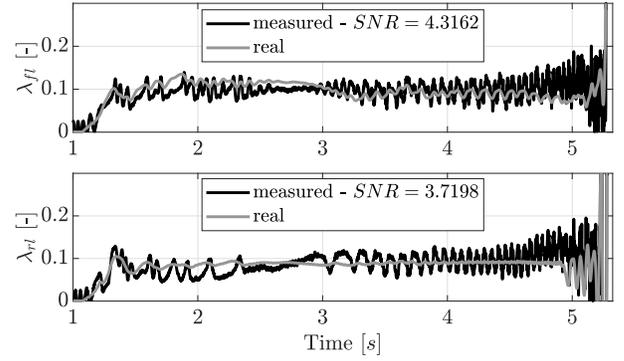}
	\caption{Noisy and noiseless slip signals, during a closed-loop braking maneuver.}
	\label{fig:snr_example}
\end{figure}

\textbf{Wheel slip.} Given the noise models defined in \eqref{eq:noise_speed} and \eqref{eq:noise_wheelrate}, the slip in \eqref{eq:slip_eq} is also affected by a noise term, depending on both noisy speed and wheel rate measurements. The noisy slip measurement $\lambda^n_{ij}$ reads
\begin{equation}
	\lambda^n_{ij}=\lambda_{ij}+n^{ij}_{\lambda}\left(v^n_{x,ij},\omega^n_{ij}\right).
\end{equation}
The terms $\sigma_{v_x}$, $k_\omega$, $\omega^n_0$ appearing in equations above are tuned to achieve signal-to-noise ratios ($SNRs$) on the slip measurements which are compatible with those observed in real setups \cite{formentin_2013}, namely $SNR\approx 4$. A comparison between noiseless and noisy slip signals is given in Fig. \ref{fig:snr_example}. 

%%%%%%%%%%%%%%%%%%%%%%%%%%%%%%%%%%%%%%%%%%%%%%%%%%%%%%%%%%%%%%%%%%%%%%%%%%%%%%%%%%%%
\section{TiL-C design}
\label{Section:silc_braking}
\begin{figure}[h]
	\centering
	\includegraphics[width=0.8 \columnwidth]{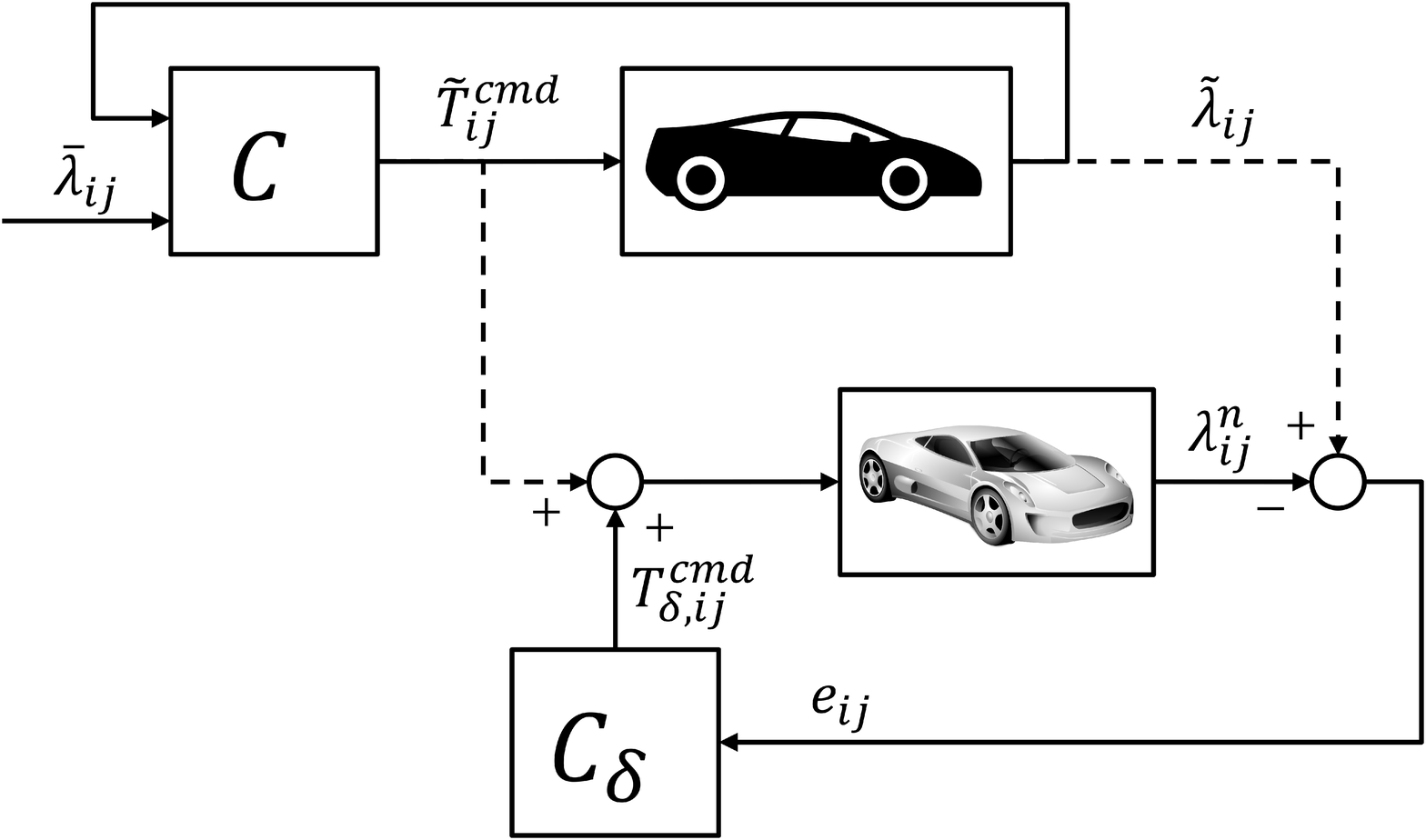}
	\caption{TiL-C braking control. The nominal MPC runs on the simulator in the above control loop, while a second loop is closed on the physical vehicle via $C_{\delta}$ to compensate for unmodelled and stochastic dynamics.}
	\label{fig:sil_c_braking}
\end{figure}

Given the vehicle model and the benchmark controller described in the previous section, we now discuss the TiL-C approach to the active braking problem, see Fig. \ref{fig:sil_c_braking}.

The nominal MPC is designed and used on the simulator and returns a torque for each wheel $\tilde{T}^{cmd}_{ij}$, corresponding to a certain (ideal) slip output $\tilde{\lambda}_{ij}$. The input to the physical vehicle is given by the sum of the nominal torque $\tilde{T}^{cmd}_{ij}$ and the outcome of the compensator $C_{\delta}$, which is non-zero anytime the measured slip is different from $\tilde{\lambda}_{ij}$.
\begin{remark}Let us point out that the TiL-C architecture proposed herein is absolutely general and independent on the employed nominal controller - \emph{i.e.} the block $C$ in Fig. \ref{fig:sil_c_braking}. For this reason, the assessment of a TiL-C scheme should be made in comparison to the nominal control loop, while the absolute performance is less relevant (in fact, the purpose of the scheme is to retrieve the desired behaviour, and not to outperform any particular strategy).
\end{remark}
\begin{remark}Note that, for an actual implementation of the controller, some activation logic for the Digital Twin is necessary: if left to freely evolve over time, the simulator might eventually diverge. For this reason, the DT has to be activated whenever the controlled maneuver starts (\emph{e.g.} when the driver starts braking). This can be simply achieved by freezing the simulation and then re-activating it by properly setting the initial states when necessary.
\end{remark}
\subsection{The TiL-C block}
\begin{figure}[h]
	\centering
	\includegraphics[width=0.9 \columnwidth]{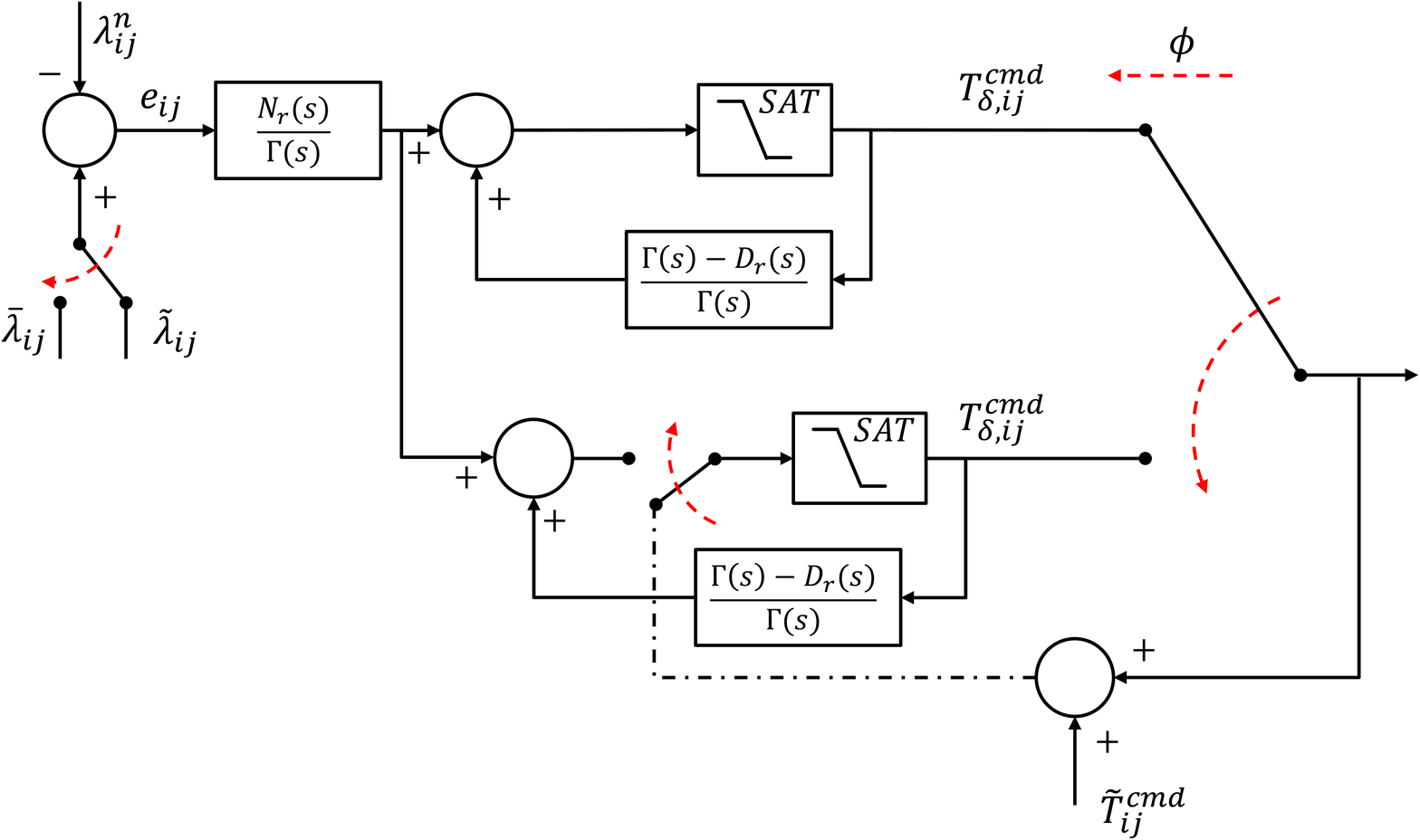}
	\caption{The detailed $C_{\delta}$ architecture.}
	\label{fig:sil_c_braking_architecture}
\end{figure}

The compensator scheme is showed in Fig. \ref{fig:sil_c_braking_architecture}. 

A linear regulator $R(s)=N_r\left(s\right)/D_r\left(s\right)$ processes the measurement error $e_{ij}$ so as to obtain the control action $T^{cmd}_{\delta,ij}$. Such a regulator is implemented in an \textit{anti-windup} fashion \cite{astrom_2006}, where a suitable de-saturation function $\Gamma\left(s\right)$ is employed in the scheme.
The regulator is selected as a Proportional-Integral (PI) one
\begin{equation}
	R\left(s\right)=\dfrac{N_r\left(s\right)}{D_r\left(s\right)}=\dfrac{k_p\left(1+sT_i\right)}{sT_i},\ \Gamma\left(s\right)=1+sT_i.
	\label{eq:PI_SIL}
\end{equation}
Then, it is discretized at sampling frequency $f_s=200\ Hz$ via the Tustin approach. Notice that the parameterization of $R\left(s\right)$ might be different for front and rear wheels, \textit{e.g.}, due to the different spin inertia or radii.

Since the wheel slip dynamics highly depend on the vehicle speed (see again \eqref{eq:predictive_slip}), we include a speed-based scheduling law, defined \textit{a-priori}, for the controller gain \cite{johansen_2003}:
\begin{equation}
	k_p=\begin{cases}
		k_p^{nom},&\ v_x\geq v_x^{ub}\\
		k_p^{nom}\cdot \left(k^{lb}_{p}+\dfrac{v_x-v_x^{lb}}{v_x^{ub}-v_x^{lb}}\right),&\ v_x^{lb}\leq v_x<v_x^{ub}\\
		k_p^{nom}\cdot k_p^{lb},&\ v_x<v_x^{lb},
	\end{cases} 
\end{equation}
%In the following, we consider the scheduling parameters to be application dependent and thus a known prior for the problem. However, tuning the controller nominal gain $k_p^{nom}$ and reset time $T_i$ is an open point.
where $v_x^{lb},\ v_x^{ub},\ k_p^{lb}$ are suitable tuning knobs.
Considering different parameters for front and rear wheels, one obtains the following parameter vector, to be tuned 
\begin{equation}
	\theta_{\textrm{TiL-C}}=\begin{bmatrix}
		k_{p,f}^{nom} & T_{i,f} & k_{p,r}^{nom} & T_{i,r}
	\end{bmatrix}.
	\label{eq:silc_parameters}
\end{equation}
%Selecting a proper value for the parameters in Eq. \eqref{eq:silc_parameters} is discussed in Section \ref{Section:ControllerTuning}.

%\subsection{Smooth-switching implementation}
%\label{Section:silc_architecture}
Notice also that, in case the simulated car completes the braking maneuver before the physical vehicle \textit{e.g.}, due to a lower simulated mass, the reference $\bar{\lambda}_{ij}$ goes to zero, and so does the nominal control $\tilde{T}^{cmd}_{ij}$.

This calls for the smooth switching structure in the lower loop of Fig. \ref{fig:sil_c_braking_architecture}, in which we introduce the signal $\phi$, driving the de-activation of the braking control for the simulated vehicle.

As soon as the simulator ends the braking maneuver earlier then the vehicle, the reference signal $\tilde{\lambda}_{ij}$ is switched to a constant value $\bar{\lambda}$. Then, a second controller tracks the total torque commanded to the system $\tilde{T}^{cmd}_{ij}+{T}^{cmd}_{\delta,ij}$. In this way, a smooth switching is guaranteed, from the full TiL control action to $C_{\delta}$ only. Such scheme \textit{de-facto} implements a soft-insertion of the control action, which is a typical solution in PI control design \cite{astrom_2006}, where $\tilde{T}^{cmd}_{ij}+{T}^{cmd}_{\delta,ij}$ serves as a manual-mode control action.
Since the accurate tuning of such a de-activation logic is not among the scopes of this research, we assume that the TiL-C is de-activated as soon as the speed hits $10\ km/h$.

\subsection{Controller tuning}
\label{Section:ControllerTuning}
As $C_{\delta}$ is aimed to control the dynamics of the residual between the system output and the output of the best available model, no model can be used to properly tune such $C_{\delta}$ in \textit{model-based fashion}. The design of this block must therefore rely \textit{only} on measurements collected on the plant.

Many data-driven methods for tuning PI controller parameters exist, see, \textit{e.g.}, \cite{yamamoto2008design,formentin2012non,formentin2014comparison,formentin2019deterministic}. However, such approaches are mainly defined for LTI systems and are not suited for the specific schemes of Fig. \ref{fig:sil_c_braking_architecture}, where an additional control action is provided, coming from the nominal closed-loop. For the above reasons, we will instead employ a Bayesian Optimization (BO) rationale \cite{khosravi_2022}, \cite{neumann_2019}. 
BO deals with black-box optimization problems, where the cost function and the constraints are unknown but the values corresponding to some instances of the decision variables can be properly ``measured''. Specifically, BO relies on the assumption that the cost function, here describing the closed-loop performance with certain control parameters, can be modeled as a Gaussian process (GP). Closed-loop data can then be used to estimate a GP model of such a cost function, while a suitable \textit{acquisition function} is chosen to select the next set of parameters to evaluate, by looking for a balanced \textit{exploration/exploitation} trade-off. For a more accurate description of the BO procedure, the interested reader can refer to, \textit{e.g.}, \cite{frazier_2018}. 

In the following, we consider BO as a decision maker, in order to solve an optimization problem of the type
\begin{equation}
	\begin{split}
		\min \ J\left(\theta\right) \\
		\textrm{subject to}\ \theta \subseteq&\Theta,
	\end{split}
	\label{eq:optimization_problem}
\end{equation}
whereas $\theta$ is a set of controller parameters, to be searched for within an \textit{a-priori} defined set $\Theta$.
Since the TiL-C goal is the control of residual dynamics between ideal and real loops, $J$ is selected as the root mean square of the average slip tracking error among the four wheels
\begin{equation}
		J\left(\theta_\textrm{{TiL-C}}\right)=\left(\sum_{k=1}^{N_s}\left(\dfrac{\sum_{ij=fl,fr,rl,rr}e^2_{ij}\left(k\right)}{4N_s}\right)\right)^{0.5}.
	\label{eq:J_lambda}
\end{equation}

where the dependence of $e_{ij},\ ij=fl,fr,rl,rr$ upon $\theta_{\textrm{TiL-C}}$ is dropped for the sake of brevity, and $N_s$ is the number of samples in the experiment. From a set of closed-loop experiments, one could re-calibrate the compensator parameters.

\begin{figure}[h]
	\centering
	\includegraphics[width=0.8 \columnwidth]{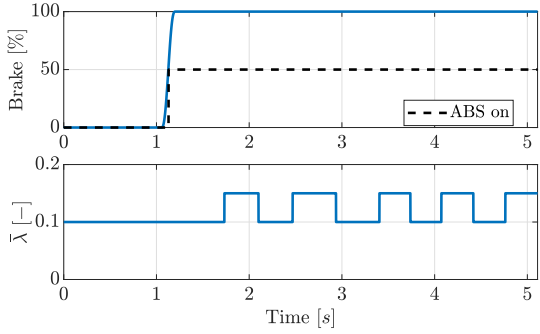}
	\caption{The training experiment for the BO procedure.}
	\label{fig:optimization_experiment}
\end{figure}

Specifically, the experiment used to train $C_{\delta}$ is illustrated in Fig. \ref{fig:optimization_experiment}. A coasting-down phase is followed by a strong braking, yielding the braking control activation. The slip reference signal is superimposed with a pulse wave varying signal, in order to better excite the system dynamics to control. 

%%%%%%%%%%%%%%%%%%%%%%%%%%%%%%%%%%%%%%%%%%%%%%%%%%%%%%%%%%%%%%%%%%%%%%%%%%%%%
\section{Simulation results}\label{Section:simulation_results}

In this section, we will show the performance of the TiL-C scheme on a test braking maneuver. As a baseline for a fair assessment of the results, we will also consider a standard EoL procedure, in which the MPC is fine tuned using the same data available for the design of $C_{\delta}$.

More specifically, since the aim of the EoL calibration is to obtain on the physical vehicle the same performance attained on the digital twin, the predictive model parameters are adjusted so as to minimize the distance between the ideal and the measured output in closed-loop. A suitable cost function to this scope is the root mean square of the wheel-averaged MPC prediction error
\begin{equation}
	\resizebox{.99\columnwidth}{!}{$
		\begin{split}
			J\left(\theta_{\textrm{MPC}}\right)&=
			&\left(\sum_{k=1}^{N_s}\left(\dfrac{\sum_{ij=fl,fr,rl,rr}\left(\lambda_{ij}\left(k\right)-\lambda^{mpc}_{ij}\left(k\right)\right)^2}{4N_s}\right)\right)^{0.5},
		\end{split}
		$}
\end{equation}
where $\lambda^{\textrm{MPC}}_{ij}$ is computed according to Eq. \eqref{eq:predictive_slip}, and projected forward in time - for each time step - depending on the prediction horizon. $\theta_{\textrm{MPC}}$ contains the predictive model parameters, found in Eq. \eqref{eq:predictive_slip}.

Three indices are employed to quantitatively assess the performance of the controllers; the first - and most important one - is the cost in Eq. \eqref{eq:J_lambda}. The second index instead represents the control effort, evaluated through the time derivative of the actuated braking torques, $\dot{T}_ij^{act}$ (\emph{e.g.} as done in \cite{tavernini_2019}) 
\begin{equation}
		J_{u}=\left(\sum_{k=1}^{N_s}\left(\dfrac{\sum_{ij=fl,fr,rl,rr}\left(\dot{T}^{act}_{ij}\left(k\right) \right)^2}{4N_s}\right)\right)^{0.5}
	\label{eq:J_u}
\end{equation}
Finally, the total braking time is displayed, defined as the time passing from the braking control activation until vehicle speed hitting $10\ km/h$.

In what follows, we will consider three case studies. In the first one, we will assume that the only difference between the digital twin and the physical vehicle is the presence of measurement noise (as expressed in Section \ref{Section:noise_model}). In the second scenario, we will neglect the effect of noise but we will investigate the case where the physical vehicle has a different mass distribution, as illustrated in Section \ref{Section:VehicleModel}. In the third scenario, we will consider both the effect of noise and the modified mass distribution, while also introducing uncertainty on the tire-road friction model, as described in Section \ref{Section:VehicleModel}. Namely, the shape factor scaling $c_s$ is set to $1.2$, while the peak friction scaling is set to $0.8$. This last test is indeed very close to what one could expect when driving an actual vehicle.
In \change{all} tests, the braking starts after $\approx 1\ s$, at the speed of $196\ km/h$. \\

All the tests in this section have been performed off-line via Matlab/Simulink environments. Real-time implementation and feasibility analysis of the controller is illustrated in Section \ref{Section:HIL}.

\subsection{The case of noisy measurements}
\begin{figure}
	\centering
	\subfloat[Commanded (solid) and actuated (dashed) torques.\label{fig:torques_noisy}]{\includegraphics[width=0.8 \columnwidth]{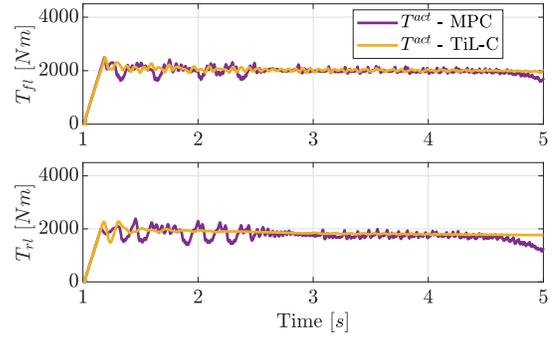}}
	\hfill
	\subfloat[Actual (solid) and reference (dashed) wheel slips.\label{fig:slips_noisy}]{\includegraphics[width=0.8 \columnwidth]{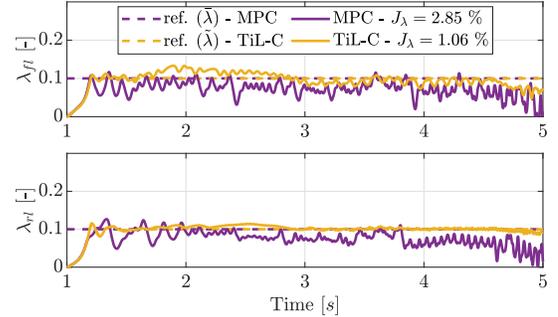}}%
	\caption{MPC and TiL-C performance in case of noisy measurements: front-left wheel (top plots) and rear-left wheel (bottom plots).}%
	\label{fig:noisy_test}%
\end{figure}

The braking experiment with noisy data is displayed in Fig. \ref{fig:noisy_test}. 
Only the front-left and the rear-left wheels are illustrated, as no significant differences between left-right corners exist during straight braking. 
\begin{figure}
		\centering
		\includegraphics[width=0.8 \columnwidth]{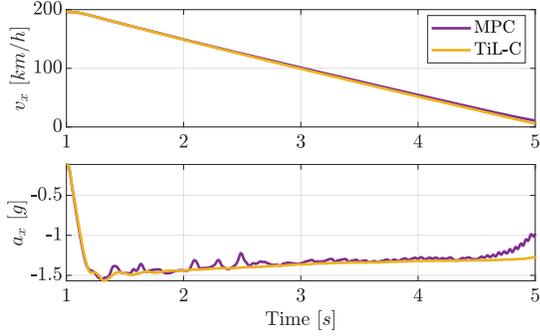}
		\caption{Vehicle speed and acceleration in case of noisy measurements, for the MPC and TiL-C controllers.}
		\label{fig:speed_noisy_SIL}
\end{figure}
Figure \ref{fig:slips_noisy} shows the reference tracking performance, whereas Figure \ref{fig:torques_noisy} depicts the actuated torques. As one can note, TiL-C is able to maintain a smoother tracking of the reference, in spite of the significant measurement noise. This is due to the fact that MPC-computed torques show undesirable oscillations induced by the presence of noise (neglected in the prediction model). 
Figure \ref{fig:speed_noisy_SIL} depicts the vehicle speed and deceleration profiles; the deceleration profile reflects what noted from the slip tracking, in that more accurate deceleration is achieved via TiL-C.
Table \ref{Tab:perf_ind_noisy} confirms what is noted in the figures: TiL-C significantly reduces both th$  $e tracking error and the control action aggressiveness, while also improving the braking time.
\begin{table}[thpb]
	\caption{Performance indices in the noisy vehicle validation experiment.}
	\label{Tab:perf_ind_noisy}
	\begin{center}	
		\begin{tabular}{cccc}
			\hline 
			& \multicolumn{3}{c}{\textbf{Performance indices}} \\
			\hline	
			& $J_{\lambda}\ [\%]$ & $J_{u}\ [Nm/s]$ & $J_{time}\ [s]$ \\ 
			\hline 
			TiL-C & $1.06$ & $4.17$ & $3.88$ \\ 
			MPC & $2.85$ & $10.82$ & $4.00$\\
			\hline
		\end{tabular}
	\end{center}
\end{table}

\begin{figure}
	\centering
	\subfloat[Commanded (solid) and actuated (dashed) torques.\label{fig:torques_perturb}]{\includegraphics[width=0.8 \columnwidth]{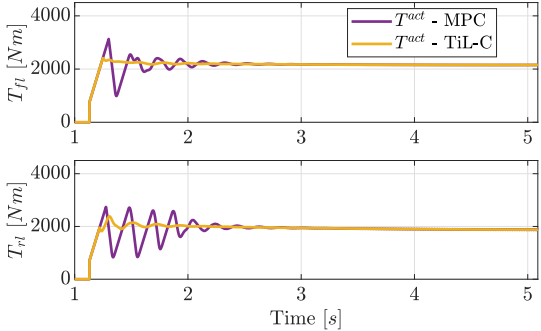}}
	\hfill
	\subfloat[Actual (solid) and reference (dashed) wheel slips.\label{fig:slips_perturb}]{\includegraphics[width=0.8 \columnwidth]{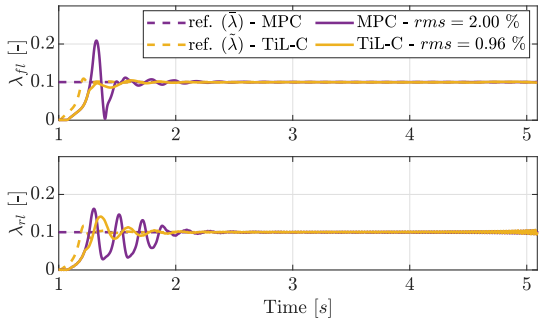}}%
	\caption{\label{fig:perturbated_test}MPC and TiL-C performance in case of different masses configuration: front-left wheel (top plots) and rear-left wheel (bottom plots).}%
\end{figure}
\begin{figure}
		\centering
		\includegraphics[width=0.8 \columnwidth]{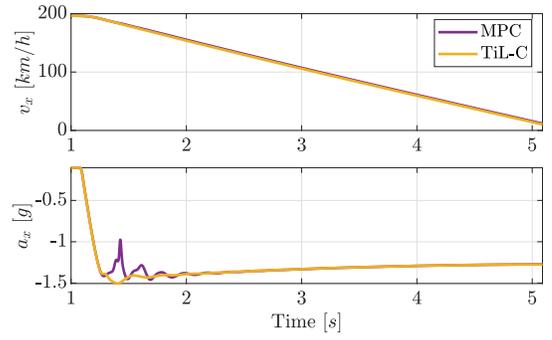}
		\caption{Vehicle speed and acceleration in case of different masses configuration, for the MPC and TiL-C controllers.}
		\label{fig:speed_perturb_SIL}
\end{figure}
\begin{figure}
	\centering
	\includegraphics[width=0.8 \columnwidth]{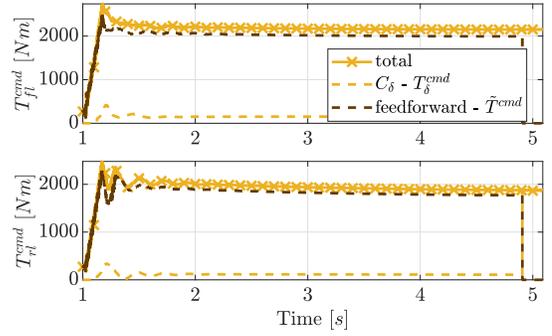}
	\caption{Split between the nominal input and the output of $C_{\delta}$, in the test with different mass configuration in the physical vehicle.}
	\label{fig:torques_perturb_SIL}
\end{figure}
\begin{figure}
	\centering
	\includegraphics[width=0.8 \columnwidth]{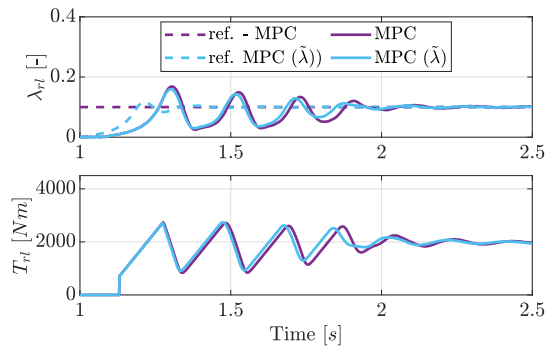}
	\caption{MPC with piecewise slip references (dark purple) and MPC with the ideal outputs as references (light blue) for the case where the physical vehicle has a different mass configuration.}
	\label{fig:mpc_comparison}
\end{figure}

\subsection{The case of additional unmodelled loads}
Let us consider the case where the physical vehicle - with noiseless data - is equipped with additional (unmodelled) masses. \\ Figure \ref{fig:perturbated_test} shows the actuated torques (upper plot) and the wheel slips (mid plot). 
As one can note, the MPC controlled vehicle exhibits significant slip tracking error and oscillations during the transient. This is due to the increased mass on the front trunk, almost leading to instability when coupled with the actuator nonlinearities. Instead, the TiL-C scheme shows good performance as the reference is well tracked with a proper torque actuation. 
Figure \ref{fig:speed_perturb_SIL} shows the vehicle speed and deceleration profiles; also in this case, the slip tracking oscillations noted in the MPC reflects into the deceleration profile, which shows oscillating behavior at the beginning of the braking maneuver.
In Figure \ref{fig:torques_perturb_SIL}, we also show the split between the contributions given by the nominal control and the TiL-C compensator. From this additional plot, it can be noted that the most significant contribution of the control action is in fact the nominal one, while $C_{\delta}$ only produces a small compensation term (thus confirming the suitability of a linear compensator for small-signal regulation).
One can also appreciate the importance of the switching architecture described in Fig. \ref{fig:sil_c_braking_architecture}. In fact, after approximately $5\ s$, the nominal torque goes to zero, as the virtual vehicle is stopping due to the reduced mass. However, the overall control action is kept at the same level due to the second loop running in parallel (Fig. \ref{fig:sil_c_braking_architecture}).

As a final remark, one might argue that feeding the MPC scheme in Fig. \ref{fig:mpc_braking} with the ideal outputs $\tilde{\lambda}_{ij}$ instead of the piecewise references might increase performance, as the former signals are smoother. However, Fig. \ref{fig:mpc_comparison} shows that no visible improvements are achieved. This is due to the fact that the traditional controller is characterized by a single block that needs to be both suitable for the nominal case and robust to parameter variations. Instead, the two blocks building the TiL-C scheme play different roles and, while the MPC is aimed only to push the nominal loop at its limits, the residual dynamics is taken care of by the additional compensator.
The above observations are confirmed by the performance indices reported in Table \ref{Tab:perf_ind_pert}.
\begin{table}[th]
	\caption{Performance indices for the case where the physical vehicle has a different mass configuration.}
	\label{Tab:perf_ind_pert}
	\begin{center}	
		\begin{tabular}{cccc}
			\hline 
			& \multicolumn{3}{c}{\textbf{Performance indices}} \\
			\hline	
			& $J_{\lambda}\ [\%]$ & $J_{u}\ [Nm/s]$ & $J_{time}\ [s]$ \\ 
			\hline 
			TiL-C & $0.96$ & $3.05$ & $3.96$ \\ 
			MPC & $2.00$ & $7.18$ & $4.01$\\
			\hline
		\end{tabular}
	\end{center}
\end{table}
For completeness, we report in Table \ref{Tab:perf_ind_pert_noisy} also a comparison between the two strategies when both measurement noises and unmodeled masses characterize the physical vehicle. As expected, the qualitative conclusions previously discussed are confirmed also in this case.
\begin{table}[th]
	\caption{Performance indices for the case where the physical vehicle has a different mass configuration and the measurements are noisy.}
	\label{Tab:perf_ind_pert_noisy}
	\begin{center}	
		\begin{tabular}{cccc}
			\hline 
			& \multicolumn{3}{c}{\textbf{Performance indices}} \\
			\hline	
			& $J_{\lambda}\ [\%]$ & $J_{u}\ [Nm/s]$ & $J_{time}\ [s]$ \\ 
			\hline 
			TiL-C & $1.84$ & $3.49$ & $4.01$ \\ 
			MPC & $3.96$ & $13.85$ & $4.19$\\
			\hline
		\end{tabular}
	\end{center}
\end{table}
\subsection{The case of modified tire-road interaction}
\begin{figure}
	\centering
	\includegraphics[width=0.8 \columnwidth]{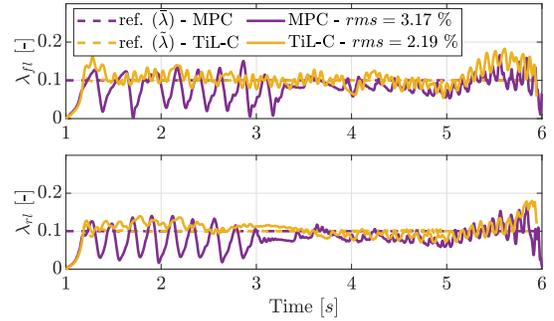}%
	\caption{\label{fig:slips_noisy_perturb_friction}MPC and TiL-C slip tracking performance in case of noisy measurements, perturbated masses and tire-road interaction: front-left wheel (top plot) and rear-left wheel (bottom plot).}%
	\label{fig:noisy_pert_fric_test}%
\end{figure}
\begin{table}[th]
	\caption{Performance indices for the case where the physical vehicle has a different mass configuration, the measurements are noisy and the tire-road interaction is modified.}
	\label{Tab:perf_ind_noisy_pert_fric}
	\begin{center}	
		\begin{tabular}{cccc}
			\hline 
			& \multicolumn{3}{c}{\textbf{Performance indices}} \\
			\hline	
			& $J_{\lambda}\ [\%]$ & $J_{u}\ [Nm/s]$ & $J_{time}\ [s]$ \\ 
			\hline 
			TiL-C & $2.19$ & $8.31$ & $4.95$ \\ 
			MPC & $3.17$ & $12.04$ & $5.00$\\
			\hline
		\end{tabular}
	\end{center}
\end{table}
Finally, we consider the case where the physical vehicle has a different mass configuration and the tire-road force generation model is modified (peak friction $\mu_s$ scaled to $0.8$, shape factor $c_s$ scaled to $1.2$). For the sake of brevity, we report here only the controller performance in terms of slip tracking, see Fig. \ref{fig:slips_noisy_perturb_friction}. As one can notice, TiL-C slightly improves the performance with respect to the nominal controller - which exhibits significant oscillations, as for the case of modified mass distribution alone of Fig. \ref{fig:slips_perturb} - and is thus robust to a low-friction test. Table \ref{Tab:perf_ind_noisy_pert_fric} displays the performance metrics; let us note that TiL-C yields a reduced control effort too. Finally, we wish to stress that the braking time is increased by the $20\ \%$ with respect to the other tests: this is explainable with the reduced maximum friction.

\section{Hardware-in-the-Loop validation}
\begin{figure*}[t]
	\centering
	\includegraphics[width=0.8 \columnwidth]{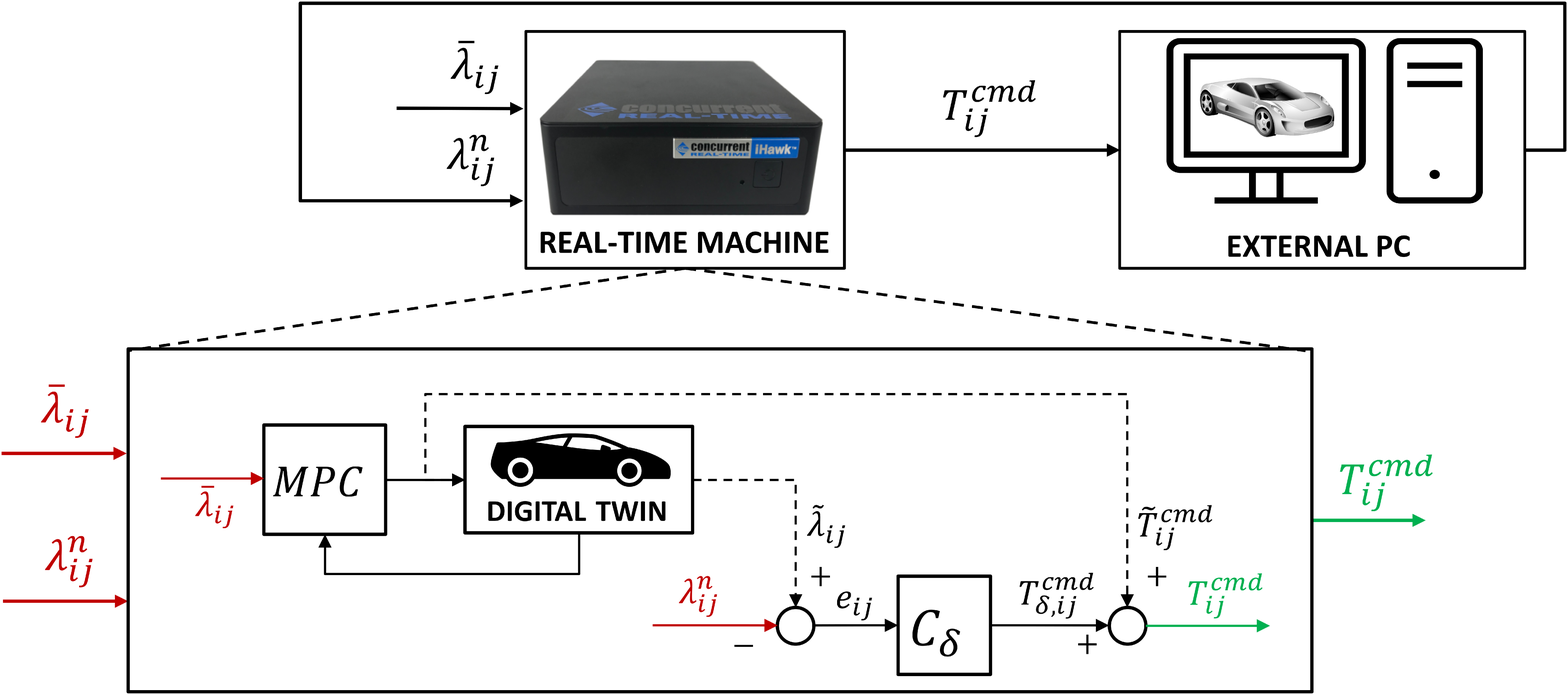}
	\caption{\label{fig:implemented_controller}Hardware-in-the-Loop experiment layout.}
\end{figure*}
\begin{figure}[th]
	\centering
	\includegraphics[width=0.8\columnwidth]{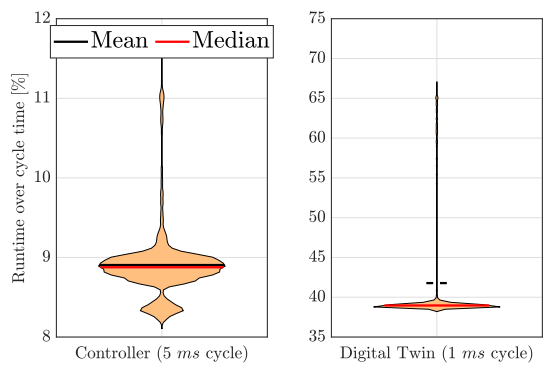}
	\caption{Summary statistics (violin plots) of the Controller (left plot) and Digital Twin (right plot) execution runtimes. The runtimes are displayed as a percentage of the corresponding cycle time.}
	\label{fig:violin_v2}
\end{figure}
\begin{figure}[th]
	\centering
	\includegraphics[width=0.8 \columnwidth]{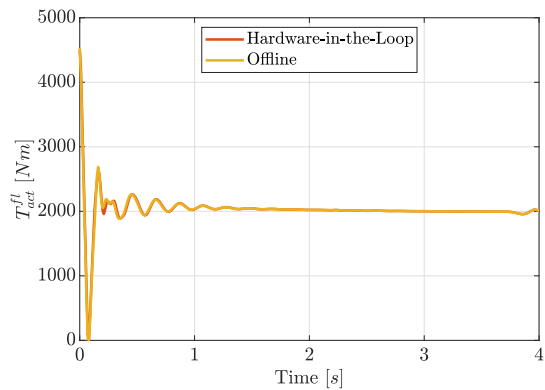}
	\caption{Time domain comparison of actuated front-left braking torque in the HiL test.}
	\label{fig:torques_comparison_HIL}
\end{figure}
\begin{figure}[th]
	\centering
	\includegraphics[width=0.8 \columnwidth]{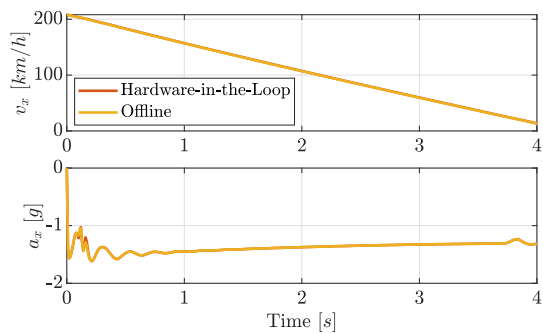}
	\caption{Time domain comparison of speed and deceleration profiles in the HIL test.}
	\label{fig:speed_acc_HIL}
\end{figure}
\label{Section:HIL}

So far, simulation analyses in an off-line fashion have been showed. In order to prove that the TiL-C framework is implementable at real-time on existing technology, we perform an HiL validation following the scheme in Fig. \ref{fig:implemented_controller}.\\
As highlighted by the figure, TiL-C is implemented in a real-time computer (RTC); more specifically, we employed AutoHawk \cite{autohawk}, an off-the-shelf Red Hat Linux based computer which is suitable for on-board usage. The available AutoHawk is provided with six $16\ GB$ RAM slots and a $3.20\ GHz$ Intel Xeon 6146 processor.\\
AutoHawk executes both the virtual and the real control loops, in details, we implement:
\begin{itemize}
	\item Four MPCs, one per each wheel, as described in Section \ref{Section:MPC}. The MPCs are implemented implicitly, \textit{i.e.} solving a QP problem at each time step, with a sampling time of $5\ ms$;
	\item Four $C_{\delta}$ compensators, one per each wheel, as described in Section \ref{Section:silc_braking}. The compensators are executed at a sampling time of $5\ ms$;
	\item The Digital Twin (modeled in CarRealTime), running at real-time with a sampling time of $1\ ms$.
\end{itemize}
An external laptop computer is solely employed to act as a vehicle simulator; the vehicle is modeled in CarRealTime and executed in Simulink (via the Real-Time toolbox). The communication among the two machines is managed by User Datagram Protocol (UDP).
The AutoHawk operating system features a scheduler which is in charge of executing the processes at real-time, according to the allowed cycle times, more specifically, two processes are executed, let apart the auxiliary ones (i.e. the scheduler itself, and the csv data logger): the first process is the Digital Twin one, the second one is the remaining part of the controller (featuring the MPC and $C_{\delta}$) The two processes are respectively executed with cycle times of $1\ ms$ and $5\ ms$.\\
In the HiL test, we consider a starting speed of $208\ km/h$, and the initial braking torque is set to its maximum value, so as to challenge the MPC and the solving time of the optimal control problem.
First, a time-domain comparison of the braking torque actuated to the front-left wheel is shown in Fig. \ref{fig:torques_comparison_HIL}: as the reader can notice, the two signals are practically superimposed - only few differences are present, due to the different solver employed by the RTC. The speed and acceleration profiles in the same test are displayed in Fig. \ref{fig:speed_acc_HIL}.\\
We are also interested in evaluating the scheduler task execution and the runtimes: Figure \ref{fig:violin_v2} shows such piece of information (a violin plot representation is employed, given that the data are not normally distributed). The left plot highlights that the execution of the controllers (the MPC and $C_{\delta}$) accounts for less than the $10\%$ of the computational load in the $5\ ms$ process: this means that significant margin exist for even more complex control logics. The right plot shows as the execution of the Digital Twin at $1\ ms$ accounts for the $\approx40 \%$ of the cycle time on average, with a long tail stretching up to the $65\%$.\\
Overall, we verified that the computationally demanding proposed strategy is feasible at real-time: we are able to execute it without any overrun, at very high sample frequencies, and employing off-the-shelf hardware.
\section{Conclusions}
\label{Section:Conclusions}
In this paper, we have proposed a new approach for vehicle dynamics control, based on the use of a full-fledged simulator, to be run \textit{in-the-loop} directly on the vehicle ECU in order to compute the nominal control action. The main advantage of such a configuration is that, in the EoL calibration phase, there is no need anymore to fine tune the (possibly many parameters of the) controller designed and calibrated on the simulator. Instead, as far as the simulator is an accurate software replica of the dynamics of the vehicle, a simple linear regulator can be used to compensate for the mismatch between the expected and the measured outputs. The new architecture shows promising results particularly when dealing with measurement noise and unmodeled terms and outperforms the state-of-the-art solution within an active braking control case study. \\
Finally, we show that the proposed architecture is real-time feasible on off-the-shelf hardware.

This being a preliminary work on the topic, many research questions remain open, \textit{e.g.}, concerning stability analysis, safe controller tuning and the generalization to different vehicle dynamics problems.

\bibliographystyle{IEEEtran}
\bibliography{ref_IEEE}
\begin{IEEEbiography}
	[{\includegraphics[width=1\columnwidth,clip,keepaspectratio]{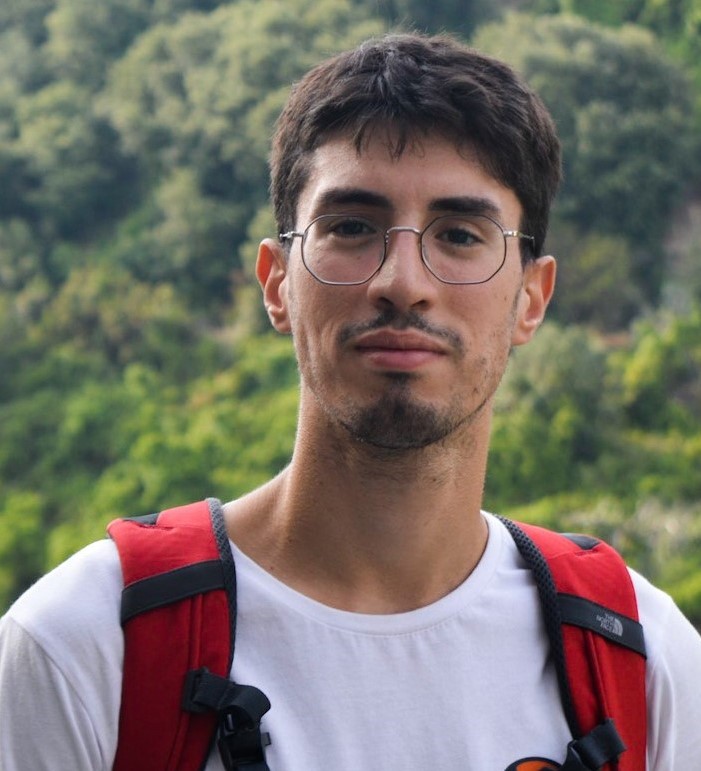}}]{Federico Dettù} received his B.Sc. on Sep. 2017 and his M.Sc. cum laude on Oct. 2020 in Automation and Control Engineering from Politecnico di Milano. From January to September 2020, he was a Visiting Researcher at Stanford University, USA.
	Starting from Nov. 2020, he joined the mOve research group as a PhD student in Information Technology, Systems and Control area, at Dipartimento di Elettronica, Informazione e Bioingegneria of Politecnico di Milano. His research interests regard data-based estimation and control approaches for automotive systems.
\end{IEEEbiography}%
\begin{IEEEbiography}
	[{\includegraphics[width=1\columnwidth,clip,keepaspectratio]{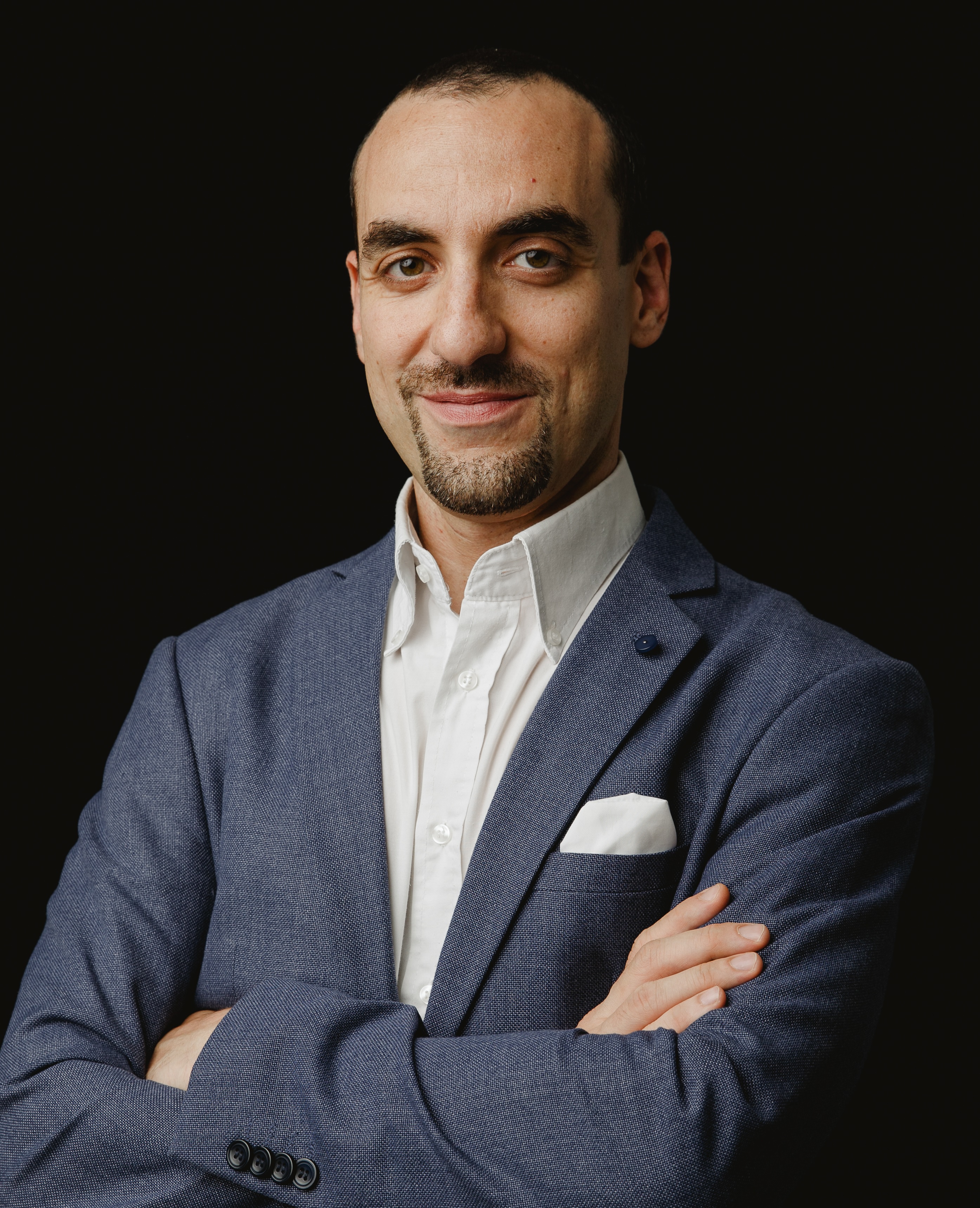}}]{Simone Formentin} was born in Legnano, Italy, in 1984. He received his B.Sc. and M.Sc. degrees cum laude in Automation and Control Engineering from Politecnico di Milano, Italy, in 2006 and 2008, respectively. In 2012, he obtained his Ph.D. degree cum laude in Information Technology within a joint program between Politecnico di Milano and Johannes Kepler University of Linz, Austria. After that, he held two postdoctoral appointments at the Swiss Federal Institute of Technology of Lausanne (EPFL), Switzerland and the University of Bergamo, Italy, respectively. Since 2014, he has been with Politecnico di Milano, first as an assistant professor, then as an associate professor. He is the chair of the IEEE TC on System Identification and Adaptive Control, the social media representative of the IFAC TC on Robust Control and a member of the IFAC TC on Modelling, Identification and Signal Processing. He is an Associate Editor of Automatica and the European Journal of Control. His research interests include system identification and data-driven control with a focus on automotive and financial applications.
\end{IEEEbiography}

\begin{IEEEbiography}
	[{\includegraphics[width=1\columnwidth,clip,keepaspectratio]{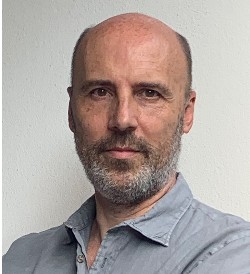}}]{Sergio Matteo Savaresi} received the M.Sc. in Electrical Engineering (Politecnico di Milano, 1992), the Ph.D. in Systems and Control Engineering (Politecnico di Milano, 1996), and the M.Sc. in Applied Mathematics (Catholic University, Brescia, 2000). After the Ph.D. he worked as management consultant at McKinsey\&Co, Milan Office. He is Full Professor in Automatic Control at Politecnico di Milano since 2006 . He is Deputy Director and Chair of the Systems\&Control Section of Department of Electronics, Computer Sciences and Bioengineering (DEIB), Politecnico di Milano. He is author of more than 500 scientific publications. His main interests are in the areas of vehicles control, automotive systems, data analysis and system identification, non-linear control theory, and control applications, with special focus on smart mobility. He has been manager and technical leader of more than 400 research projects in cooperation with private companies. He is co-founder of 8 high-tech startup companies.
\end{IEEEbiography}

\end{document}